\documentclass[12pt]{article}

\usepackage{amsmath,amssymb,amsthm, bbm}
\usepackage{hyperref}
\usepackage{graphicx}
\usepackage{natbib}
\usepackage{authblk}
\usepackage{xcolor}
\usepackage[ruled]{algorithm2e}

\usepackage{tcolorbox}
\usepackage{booktabs,multirow}
\usepackage{adjustbox}
\newcommand{\blind}{1} 

\newtheorem{theorem}{Theorem}

\newtheorem{proposition}{Proposition}

\newtheorem{remark}{Remark}

\def\I{{\bf I}}

\def\R{{\bf R}}

\def\T{{\bf T}}

\def\X{{\bf X}}
\def\x{{\bf x}}

\def\bX{{\bf X}}

\def\bb{{\boldsymbol\beta}}

\def\bt{{\boldsymbol\theta}}
\def\bT{{\boldsymbol\Theta}}

\def\bmu{\boldsymbol\mu}

\def\bSigma{{\boldsymbol\Sigma}}

\def\ps{p_s}
\def\pt{p_t}

\def\0{{\bf 0}}
\def\1{{\bf 1}}
\def\trans{^{\rm T}}

\def\wh{\widehat}
\def\wt{\widetilde}

\def\log{{\rm log}}

\def\bbE{\mathbb{E}}

\def\red{\textcolor{black}}
\def\blue{\textcolor{black}}

\newcommand{\nc}{\newcommand}
\nc{\nn}{\noindent}
\definecolor{watpink}{RGB}{198, 0, 120}
\definecolor{watyellow}{RGB}{255, 213, 79}
\definecolor{skyblue}{RGB}{0, 102, 204}
\nc{\green}[1]{\textcolor{green}{#1}}
\nc{\purple}[1]{\textcolor{purple}{#1}}
\nc{\violet}[1]{\textcolor{violet}{#1}}
\nc{\uwpink}[1]{\textcolor{watpink}{#1}}
\nc{\bI}{\boldsymbol I}
\nc{\bbeta}{\boldsymbol \beta}
\nc{\cmy}[1]{{\color{watpink}{[YCK Comment: #1]}}}
\nc{\yck}[1]{{\color{watpink}{#1}}}

\definecolor{jade}{rgb}{0.0, 0.66, 0.42}
\nc{\update}[1]{\textcolor{black}{#1}}

\usepackage[normalem]{ulem}
\nc{\co}[1]{\sout{\red{#1}}}

\newcommand{\mytitle}{Positive and Unlabeled Data: Model, Estimation, Inference, and Classification}
\title{\LARGE\bf\mytitle}
\author[1,2]{Siyan Liu}
\author[2]{Chi-Kuang Yeh}
\author[3]{Xin Zhang}
\author[2]{Qinglong Tian}
\author[2]{Pengfei Li}
\affil[1]{KLATASDS-MOE, School of Statistics, East China Normal University}
\affil[2]{Department of Statistics and Actuarial Science, University of Waterloo}
\affil[3]{Department of Statistics, Iowa State University}
\date{     \begin{center}
        {\small
        \today}
    \end{center}
}

\begin{document}
	
	\def\spacingset#1{\renewcommand{\baselinestretch}%
		{#1}\small\normalsize} \spacingset{1}
	
	\if1\blind
	{
		\maketitle
	} \fi
	
	\if0\blind
	{
		\bigskip
		\bigskip
		\bigskip
       
		\begin{center}
			{\LARGE\bf\mytitle}
		\end{center}
		\medskip
	}\fi
	
	\bigskip

\begin{abstract}
This study introduces a new approach to addressing the positive and unlabeled (PU) data through the double exponential tilting model (DETM) under a transfer learning framework.
Traditional methods often fall short because they only apply to the \blue{common distributions (CD)} PU data (also known as the selected completely at random PU data), where the labeled positive and unlabeled positive data are assumed to be from the same distribution.
In contrast, our DETM's dual structure effectively accommodates the more complex and underexplored \blue{different distribution (DD)} PU data (also known as the selected at random PU data), where the labeled and unlabeled positive data can be from different distributions.
We rigorously establish the theoretical foundations of DETM, including identifiability, parameter estimation, and asymptotic properties.
Additionally, we move forward to statistical inference by developing a goodness-of-fit test for the \blue{CD} assumption and constructing confidence intervals for the proportion of positive instances in the target domain.
We leverage an approximated Bayes classifier for classification tasks, demonstrating DETM's robust performance in prediction.
Through theoretical insights and practical applications, this study highlights DETM as a comprehensive framework for addressing the challenges of PU data.
\end{abstract}
	
	\noindent%
	\spacingset{1.9} 

\section{Introduction}

\subsection{Positive and Unlabeled Data}

Considering a binary response variable $Y\in\{0,1\}$ and features $\X\in\mathbb{R}^{p}$, positive and unlabeled (PU) data refer to datasets that consist of the following two samples:
The first sample is labeled, but all observations in this sample have positive responses $y=1$;
the second sample contains both positive ($y=1$) and negative ($y=0$) instances.
However, the second sample is unlabeled, meaning we can only access the features $\x$ but have no access to their labels $y$.
{PU} data are prevalent in many areas, including biomedical study, ecology, and machine learning.

\paragraph{Contaminated Case-Control Data}

A case-control study is a research design widely used in biomedical research to investigate the causes of diseases.
It compares two groups: a group of patients with the disease (the case group) and a group of healthy people (the control group).
Data contamination is a common issue in case-control studies.
For example, the underdiagnosis phenomenon refers to a situation where the false positive rate is very low, and the case group can generally be seen as pure and only contains patients.
However, some medical conditions may go unrecognized in many patients, and they are misclassified as healthy ones.
When underdiagnosis happens, many people in the control group are, in fact, patients.
For example, \citet{begg1994methodological} and \citet{godley1999adjusted} highlight an issue in a prostate cancer study, indicating that 20\%-40\% of the control group consists of latent prostate cancer cases, rendering the control group contaminated.
In the underdiagnosis example, the case group can be seen as a positive-only sample, but the contaminated control group is an unlabeled sample containing both positive and negative cases because we do not know who are real healthy people and patients.
\update{
In addition to biomedical studies, contaminated case-control data are also prevalent in econometrics (e.g., \citealt{LANCASTER1996145}).
}

\paragraph{Presence-Only Data}

Presence-only data can come from various fields, including ecology, epidemiology, and marketing.
For example, ecological studies often rely on presence-only data to model a species' habitat (\citealt{ward2009presence}).
The dataset comprises two samples: one sample contains confirmed presence locations of a species obtained through field surveys or other methods; the other sample contains some other locations sampled from the full study area.
However, the presence status of the species in the latter sample is unknown.
Thus, the first sample is positive only while the second sample contains both presence and non-presence locations but is unlabeled, meaning that we do not know which location has the confirmed presence of the species.

\paragraph{PU Data in Machine Learning}

PU data are commonly seen in many machine learning tasks.
For instance, on an online video streaming platform, users click on videos that they find interesting.
However, for videos that are not clicked, it is uncertain whether users are interested in them or not, as users may ignore those videos while browsing.
In such a scenario, those clicked videos are positive-only data, while unclicked videos are considered unlabeled data for a recommendation system.
In computer vision, tasks like facial recognition are often based on PU data.
Take this specific example: We aim to identify a user from various images using some known reference pictures of that person.
Here, the labeled sample comprises all clearly identified face images of that user.
The unlabeled sample we wish to identify consists of photos of the user and other people.
We refer interested readers to read \citet{jaskie_positive_2019} for more examples of PU data in machine learning.

\subsection{Notation and Motivations}

Using the terminologies from transfer learning, we consider a source domain $\ps(\x,y)$ and a target domain $\pt(\x,y)$.
In the rest of this paper, we use the generic notation $p_s(\cdot)$ (and $\Pr_s(\cdot)$) and $p_t(\cdot)$ (and $\Pr_t(\cdot)$) to denote densities (and probabilities) on the source and target domains, respectively.
The labeled positive data are sampled from the source domain
\[
{\left\{(\x_i,y_i=1),i=1,\dots,n\right\}\sim \ps(\x|y=1).}
\]
The unlabeled data are from the target domain
\begin{equation}
\label{eq:unlabeled-data-mixture}
\left\{\x_i,i=n+1,\dots,n+m\right\}\sim \pt(\x).
\end{equation}
The marginal distribution of $\X$ in (\ref{eq:unlabeled-data-mixture}) can be seen as a mixture distribution
\[
\pt(\x)=\pi\pt(\x|y=1)+(1-\pi)\pt(\x|y=0),
\]
where $\pi$ is the mixture proportion $\pi\equiv\Pr_t(y=1)$ of positive cases in the target domain.

Even though much research has been done on PU data, most of it suffers from two limitations.
The first limitation is that many existing works require the mixture proportion $\pi$ to be known.
However, such an assumption is usually unrealistic as $\pi$ is rarely known.

The second and major limitation is that most existing research works are constrained to the \blue{common distributions (CD)} PU data (also known as the selected at random PU data, see \citealt{bekker_learning_2020}), which imposes the assumption that $p_s(\x|y=1)=p_t(\x|y=1)$.
The \blue{CD} assumption has been widely used for simplicity: one of the mixture components, $p_t(\x|y=1)$, is readily identifiable because the positive-only data $\left\{(\x_i,y_i=1),i=1,\dots,n\right\}$ are from the same distribution as $p_t(\x|y=1)$.
Despite its popularity, the \blue{CD} assumption does not always reflect the true underlying data-generating mechanism.
{In the contaminated case-control study example}, it is reasonable to believe that the latent prostate cancer
cases were misdiagnosed as healthy ones not only because of bad luck but also because their disease
symptoms were not as severe (or typical) as patients in the case group (i.e., $p_s(\x|y=1)\neq p_t(\x|y=1)$).

\begin{figure}[!ht]
    \centering
    \includegraphics[width=0.72\linewidth]{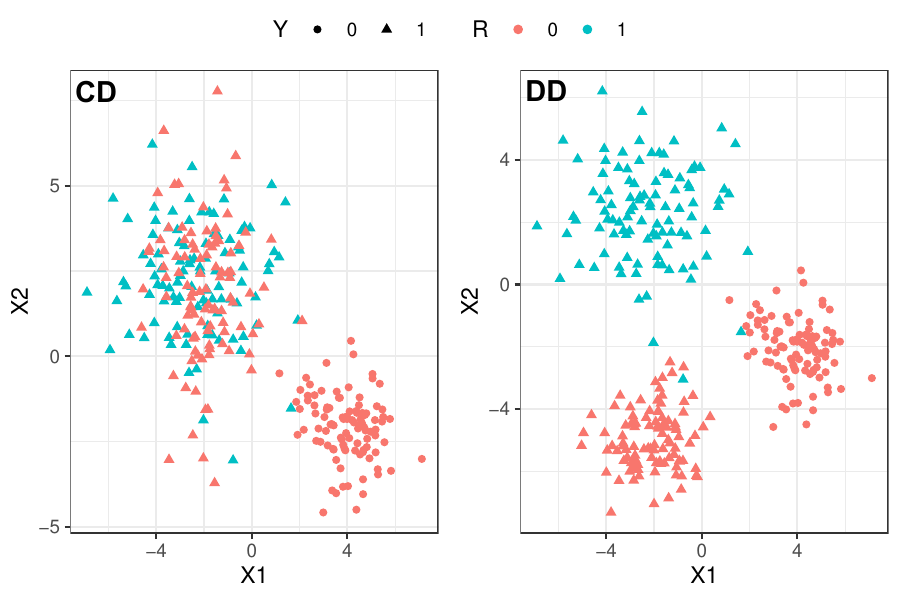}
    \caption{
    We use $R=0$ (red) and $R=1$ (blue), respectively, to denote data from the target and source.
    In the \blue{CD} setting, the labeled positive (\textcolor{black}{blue} triangles) and unlabeled positive (\textcolor{black}{red} triangles) are from the same distribution.
    However, they may be different in the \blue{DD} setting.}
    \label{fig:scar_sar-1}
\end{figure}

As opposed to the \blue{CD} assumption, the \blue{different distributions (DD)} PU setting (also known as selected-at-random PU, see \citealt{bekker_learning_2020}) allows $p_s(\x|y=1)$ to be different from $p_t(\x|y=1)$, thus it can better reflect the true data-generating mechanism.
Figure~\ref{fig:scar_sar-1} shows the difference between \blue{DD} and \blue{CD} PU data.
However, despite its flexibility, the \blue{DD} PU setting presents numerous challenges, the foremost being the identifiability of the distributions of unlabeled positive and unlabeled negative data (i.e., $p_t(\x|y=1)$ and $p_t(\x|y=0)$).

\blue{We aim to address the problem of statistical inference and classification on PU data by adopting the DD setting, which relaxes the often unrealistic CD assumption. A key advantage of our approach is that it does not require prior knowledge of the mixture proportion $\pi$.
}

\subsection{Related Work and Challenges}

PU data have been studied under different names in the statistical literature, including contaminated case-control data (e.g., \citealt{qin1994empirical, LANCASTER1996145, qin1999empirical, qin2011hypothesis, duan2020fast}) and presence-only data (e.g., \citealt{ward2009presence, song2019pulasso}).
However, the aforementioned works are constrained under the CD assumption and often rely on knowing the true value of the mixture proportion.
One exception is \citet{wang2021case}, where they consider the  DD setting.
However, they further require a smaller labeled verification sample from the target distribution.
Thus, data from their setting cannot be seen as PU data.

In machine learning literature, most works on PU data focus on the \blue{CD} assumption (e.g., \citealt{elkan_learning_2008, ramaswamy_mixture_2016, garg_mixture_2021}).
However, a few works try to address the \blue{DD} setting.
\citet{kato2018learning} impose an ``invariance of order'' assumption, which states that, for observations in the labeled source data, the higher the probability $\Pr_s(y=1|\x)$ is, the closer it is to {the unlabeled positive distribution} $p_t(\x|y=1)$.
In addition, their method requires the mixture proportion $\pi=\Pr_t(y=1)$ to be known.
\citet{bekker2020beyond} propose a method similar to inverse probability weighting, which requires estimating the propensity score function.
To ensure identifiability, they assume that the propensity score function needs to depend on fewer features than the predictive model (i.e., the shadow variable assumption in the missing data literature).
\citet{coudray2023risk} further establish risk bounds based on \citet{bekker2020beyond}.
Nevertheless, the shadow variables are generally unknown, let alone their existence, thus reducing the method's applicability.
\citet{gong_instance-dependent_2021} follow a similar framework as in \citet{bekker2020beyond} but do not impose the shadow variable assumption.
However, they require a highly restrictive assumption that the propensity score function has a value of less than 0.5 for all observations from the unlabeled data under the logistic regression model.
\citet{furmanczyk_joint_2022} use a similar idea as \citet{bekker2020beyond} and \citet{gong_instance-dependent_2021}, in which they prove the model identifiability under logistic regression.
However, their prediction approach requires that the positive and negative instances in the unlabeled data are separable.
In addition, existing works on the \blue{DD} PU setting focus solely on classification (i.e., prediction) and do not address any statistical inference problems.
Problems like the the goodness-of-fit of the \blue{CD} condition, and inference of the mixture proportion $\pi$ are remained unexplored.

\subsection{Our Contributions and Overview}

\blue{
Unlike the traditional propensity score framework used in existing literature (\citealt{bekker2020beyond, bekker_learning_2020}), we treat the PU data based on a transfer learning framework where the data consist of two samples (a detailed comparison of the two frameworks is provided in Section~\ref{sec-conclu}).
}
We further propose a new model, called the double exponential tilting model (DETM), for the DD PU data, and prove its identifiability.
Based on the proposed model, we develop an Expectation-Maximization (EM) algorithm for estimating parameters using the empirical likelihood (EL) method.
It is worth noting that our method works even if the mixture proportion $\pi$ is unknown.
Lastly, we apply the proposed model 
to 
address inference and prediction tasks, including model selection, confidence intervals, hypothesis testing, and classification. 
The highlights of our contributions are summarized in the following.
\begin{itemize}
\item Our proposed DETM, based on modeling the density ratios, is identifiable up to label switching.
With some further mild assumptions, the model becomes fully identifiable.
\item We propose a simple EM algorithm for estimating parameters, including the mixture proportion and odds ratios.
\item We investigate the asymptotic behavior of our proposed estimator and propose hypothesis testing procedures for the mixture proportion and the goodness-of-fit of the CD condition.
\item We use an approximated Bayes classifier based on the proposed model for classification tasks on the target domain.
\item We implement the proposed methods in a R package, \texttt{PUEM}, which is available in the supplementary materials. 
\end{itemize}

The rest of this paper is organized as follows.
Section~\ref{sec:model-detm} introduces the proposed DETM.
Section~\ref{sec:detm} discusses how to estimate parameters under the proposed model.
Section~\ref{sec:theory} establishes theoretical results for statistical inference and classification.
Section~\ref{sec:numerical} examines the empirical performance of the proposed methods using simulation studies and a real data application.
We close the paper with concluding remarks given in Section \ref{sec-conclu}.
All the technical details and proofs are given in the supplementary materials.

\section{Model}
\label{sec:model-detm}
\subsection{Density Ratio is the Key}
\label{subsec:motivation-det}

A commonly used strategy in distributional shift problems (or, broadly speaking, transfer learning) for connecting the source and target domains is to utilize the density ratio.
For example, adjusting the covariate shift often relies on estimating the density ratio of the features $\x$ between the target and source domains (e.g., \citealt{shimodaira2000improving, sugiyama2007covariate}), and adjusting the label shift often requires estimating the density ratio of the response between the target and source domains (e.g., \citealt{lipton_detecting_2018, tian2023elsa, kim2024retasa}).

The aforementioned works on transfer learning suggest that modeling the density ratios is a promising direction for solving the PU problem, which also involves a source and a target domain and belongs to the broad concept of transfer learning.
In fact, examining the Bayes classifier for the target domain also reveals the key role of the density ratios.
The Bayes classifier utilizes the posterior probability of the label and is optimal because it minimizes the prediction risk.
The Bayes classifier on the target domain is defined as
\[
\mathcal{C}(\x)=\mathbbm{1}\left\{{\Pr}_t(Y=1|\X=\x)>0.5\right\},
\]
where $\mathbbm{1}(\cdot)$ is the indicator function.
We can rewrite ${\Pr}_t(Y=1|\X=\x)$ using the Bayes rule as
\[
{\Pr}_{t}(Y=1|\X=\x)=\frac{\pt(\x,y=1)}{\pt(\x)}=\frac{\pi\pt(\x|y=1)}{\pi\pt(\x|y=1)+(1-\pi)\pt(\x|y=0)}.
\]
We can further write ${\Pr}_{t}(Y=1|\X=\x)$ by dividing both the numerator and denominator by $\ps(\x|y=1)$:
\begin{equation}
\label{eq:bayes-posterior}
{\Pr}_{t}(Y=1|\X=\x)=\frac{\pi\Lambda_1(\x)}{\pi\Lambda_1(\x)+(1-\pi)\Lambda_2(\x)},
\end{equation}
where $\Lambda_1(\x)$ and $\Lambda_2(\x)$ are density ratios and are defined as
\[
\Lambda_1(\x)\equiv\frac{\pt(\x|y=1)}{\ps(\x|y=1)}~~\text{ and }~~\Lambda_2(\x)\equiv\frac{\pt(\x|y=0)}{\ps(\x|y=1)}.
\]
Equation~(\ref{eq:bayes-posterior}) implies that the Bayes classifier hinges on the mixture proportion $\pi$ and two density ratios $\Lambda_1(\x)$ and $\Lambda_2(\x)$.
This finding also points out that it is promising to model the DD PU data using the density ratios.

\subsection{Identifiability Problem}
\label{subsec:identifiability-problem}
We have underscored the importance of $\Lambda_1(\x)$ and $\Lambda_2(\x)$ in the previous section.
In this section, we will show that if $\Lambda_1(\x)$ and $\Lambda_2(\x)$ are unconstrained, $\pi$, $\Lambda_1(\x)$, and $\Lambda_2(\x)$ are non-identifiable.

The unlabeled data are from the distribution $\pt(\x)$, which can be rewritten as
\begin{equation}
\label{eq:mixture-identifiability}
\begin{split}
\pt(\x)=&\pi\pt(\x|y=1)+(1-\pi)\pt(\x|y=0)\\
=&\ps(\x|y=1)\left\{\pi\Lambda_1(\x)+(1-\pi)\Lambda_2(\x)\right\}.
\end{split}
\end{equation}
The left side of (\ref{eq:mixture-identifiability}) (i.e., $\pt(\x)$) is identifiable because we have the unlabeled target data $\left\{\x_i,i=n+1,\dots,n+m\right\}$.
For the right side, $\ps(\x|y=1)$ is also identifiable because we have access to the labeled source data $\left\{(\x_i,y_i=1),i=1,\dots,n\right\}$.
Thus, the model identifiability depends on $\pi\Lambda_1(\x)+(1-\pi)\Lambda_2(\x)$.

However, for any $\pi^\ast\in(0,1)$ that satisfies $\pi^\ast<\pi$, we have the following equation
\[
\pi\Lambda_1(\x)+(1-\pi)\Lambda_2(\x)=\pi^\ast\Lambda_1(\x)+(1-\pi^\ast)\Lambda^\ast_2(\x),
\]
where
\[
\Lambda_2^\ast(\x)\equiv\frac{\pi-\pi^\ast}{1-\pi^\ast}\Lambda_1(\x)+\frac{1-\pi}{1-\pi^\ast}\Lambda_2(\x).
\]
Here $\Lambda_2^\ast(\x)$ is also a density ratio because it satisfies $\int\Lambda_2^\ast(\x)\ps(\x|y=1)d\x=1$ and $\Lambda^\ast_2(\x)\geq0$.
Therefore, both $\left\{\pi^\ast,\Lambda_1(\x),\Lambda_2^\ast(\x)\right\}$ and $\left\{\pi,\Lambda_1(\x),\Lambda_2(\x)\right\}$ satisfy (\ref{eq:mixture-identifiability}).
In other words, $\pi$, $\Lambda_1(\x)$, and $\Lambda_2(\x)$ are non-identifiable if we do not further impose more assumptions.

\subsection{Double Exponential Tilting Model} 

Previous discussions reveal that we must impose some constraints on $\Lambda_1(\x)$ and $\Lambda_2(\x)$ to solve the identifiability issue.
The exponential tilting model (ETM) is a popular choice for modeling the density ratios.
For example, \citet{maity2023understanding} used the ETM to model the density ratio between the source and target domains in a transfer learning setting.
Also, the ETM is widely used in areas such as case-control studies (e.g., \citealt{qin1998inference, qin2011hypothesis, qin_biased_2017}).

Based on previous discussions, we use the ETM to connect the source and target domains.
Rather than using an exponential tilting model for $p_t(\x,y)/p_s(\x,y)$ (e.g., \citealt{maity2023understanding}), we acknowledge that the target distribution $p_t(\x)$ is a mixture of $p_t(\x|y=1)$ and $p_t(\x|y=0)$.
Thus, we connect the source sample distribution $p_s(\x|y=1)$ with the two mixture components $p_t(\x|y=1)$ and $p_t(\x|y=0)$ using two ETMs separately.
Thus, we name the proposed model as the DETM, where the details are given below.

We impose {the ETM} to both $\Lambda_1(\x)$ and $\Lambda_2(\x)$ as follows.
\begin{equation}
\label{eq:detm}
\log\left\{\Lambda_1(\x)\right\}=\alpha_1+\x\trans\bb_1,\quad\log\left\{\Lambda_2(\x)\right\}=\alpha_2+\x\trans \bb_2,
\end{equation}
where $\Lambda_1(\x)$ and $\Lambda_2(\x)$ are given by
\[
\Lambda_1(\x)=\frac{\pt(\x|y=1)}{\ps(\x|y=1)}~~
\text{ and }
~~
\Lambda_2(\x)=\frac{\pt(\x|y=0)}{\ps(\x|y=1)}.
\]
In the DETM in (\ref{eq:detm}),  $\alpha_1$ and $\alpha_2$ are two normalizing constants such that the two density ratios must satisfy
\begin{equation}
\label{eq:normalizing-constant}
\int \Lambda_1(\x)\ps(\x|y=1)d\x=1~~
\text{ and }
~~
\int\Lambda_2(\x)\ps(\x|y=1)d\x=1.
\end{equation}

We use a linear combination $\x\trans \bb$ in (\ref{eq:detm}), and one can generally replace $\x$ with a known function of $\T(\x)$.
For example, we can use a polynomial function $\T(x)=(x,x^2,x^3)$ to increase the model's flexibility.
For high-dimensional features such as images, we can use the embedding layer of a pre-trained neural network as $\T(\x)$.
We will continue to use $\x\trans \bb$ to facilitate presentations in the rest of the paper.

The remaining question is to address the identifiability of the DETM.
As we have shown in (\ref{eq:mixture-identifiability}), the key to model identifiability lies in $\pi\Lambda_1(\x)+(1-\pi)\Lambda_2(\x)$.
In the next theorem, we prove that $\pi\Lambda_1(\x)+(1-\pi)\Lambda_2(\x)$ is ``almost'' identifiable and can become fully identifiable \update{under some non-restrictive conditions}. 
The proof is given in the supplementary materials.

\begin{theorem}[Identifiability]
\label{theorem:identifiability}
Under the DETM in \eqref{eq:detm}, we model the density ratios as $\Lambda_1(\x)=\exp\left(\alpha_1+\x\trans\bb_{{1}}\right)$ and $\Lambda_2(\x)=\exp(\alpha_2+\x\trans\bb_ {{2}})$, and we let $\pi={\Pr}_{t}(Y=1)$.
If $(\pi,\alpha_1,\bb_1,\alpha_2,\bb_2)$ and $(\pi^\ast,\alpha^\ast_1,\bb^\ast_1,\alpha^\ast_2,\bb^\ast_2)$ satisfy
\[
\pi\exp\left(\alpha_1+\x\trans\bb_1\right)+(1-\pi)\exp\left(\alpha_2+\x\trans\bb_2\right)=\pi^\ast\exp\left(\alpha_1^\ast+\x\trans\bb_1^\ast\right)+(1-\pi^\ast)\exp\left(\alpha_2^\ast+\x\trans\bb_2^\ast\right),
\]
for any $\x\in\mathbb{R}^p$, then the two sets of parameters are either the same
(i.e., $
(\pi,\alpha_1,\bb_1,\alpha_2,\bb_2)=(\pi^\ast,\alpha^\ast_1,\bb^\ast_1,\alpha^\ast_2,\bb^\ast_2)
$) or subject to label switching: $(\pi,\alpha_1,\bb_1,\alpha_2,\bb_2)=(1-\pi^\ast,\alpha^\ast_2,\bb^\ast_2,\alpha^\ast_1,\bb^\ast_1)$.
\end{theorem}
\begin{remark}
Theorem~\ref{theorem:identifiability} reveals that the DETM is identifiable up to exchanging the label $Y$.
The label switching issue is inherent in the model specification in (\ref{eq:detm}) due to the symmetry of $\Lambda_1(\x)$ and $\Lambda_2(\x)$.
We can achieve full model identifiability by introducing some reasonable assumptions to solve the label-switching problem.
For instance, if positive cases are less frequent in the unlabeled dataset (e.g., misdiagnosis is rare), we can impose the condition $\pi < 0.5$, which immediately resolves the label-switching problem.
In the rest of this paper, we assume the DETM is fully identifiable.
\end{remark}

\begin{remark}
The proposed DETM can be applied to both CD and DD PU data because CD data can be viewed as a special case of DD data with additional constraints. 
Under the CD assumption, the distribution of features
of positive cases remains the same in the source and target domains.
That is
\begin{equation}
	\label{eq:setm}
\Lambda_1(\x)
=\frac{\pt(\x|y=1)}{\ps(\x|y=1)}=1,
\end{equation}
for any $\x$.
We refer to the constrained model for CD PU data as the single exponential tilting model (SETM) because we only have the following model assumption:
\[
\log\left\{\Lambda_2(\x)\right\}=\alpha_2+\x\trans\bb_2.
\]
The SETM is a reduced model of DETM because we consider a reduced parameter space where $\alpha_1=0$ and $\bb_1=\0$ (implied by (\ref{eq:setm})).
The SETM is similar to the method used in \citet{LANCASTER1996145} in spirit and can also be seen as an extension and refinement of the methods  in \citet{qin1999empirical} and  \citet{qin2011hypothesis} as these methods are constrained to scalar $x$.
\end{remark}
{
\color{black}
\begin{remark}
The DETM in (\ref{eq:detm}) inherently assumes that
\( p_s(\mathbf{x} \mid y=1) \) dominates both \( p_t(\mathbf{x} \mid y=1) \) and \( p_t(\mathbf{x} \mid y=0) \), with all three distributions sharing a common support. In other words, this requires \( \Lambda_1(\mathbf{x}) > 0 \) and \( \Lambda_2(\mathbf{x}) > 0 \) on the common support. For instance, if \( p_t(\mathbf{x} \mid y=0) \) and \( p_s(\mathbf{x} \mid y=1) \) are perfectly separable, the assumption \( \Lambda_2(\mathbf{x}) > 0 \) will not hold, rendering our method inapplicable.  
In practice, the dominance assumption implies that data from \( p_t(\mathbf{x} \mid y=0) \) (and similarly, data from \( p_t(\mathbf{x} \mid y=1) \)) should preferably overlap to some extent with data from \( p_s(\mathbf{x} \mid y=1) \). In contrast, some methods rely on the separability assumption (e.g., \citealt{garg_domain_2022}), require that the data are well-separated and may fail when significant overlap exists.  
It is important to emphasize that dealing with overlapping data is, in fact, more challenging. When \( p_t(\mathbf{x} \mid y=1) \) is well-separated from other data, identifying examples from \( p_t(\mathbf{x} \mid y=1) \) becomes straightforward. However, in overlapping scenarios, distinguishing between classes is considerably more difficult, highlighting the practical relevance of our approach.  \end{remark}
}

\section{Parameter Estimation}
\label{sec:detm}

\subsection{Empirical Likelihood Method}

Based on the DETM, after combining the source and target data, the log-likelihood function is given by

\begin{equation}
\label{eq:log-likelihood-function}
\begin{split}
\ell&=\sum_{i=1}^{n}\log\left\{\ps(\x_i|y=1)\right\}+\sum_{j=1}^{m}\log\left\{\pt(\x_{n+j})\right\}\\
&=\sum_{i=1}^{N}\log\left\{\ps(\x_i|y=1)\right\}+\sum_{j=1}^{m}\log\left\{\pi\exp(\alpha_1+\x_{n+j}\trans\bb_1)+(1-\pi)\exp(\alpha_2+\x_{n+j}\trans\bb_2)\right\},
\end{split}
\end{equation}
where $N\equiv n+m$.
In addition, $(\alpha_1,\alpha_2,\bb_1,\bb_2)$ and $\ps(\x|y=1)$ must satisfy the conditions in (\ref{eq:normalizing-constant}), which can be rewritten as
\begin{equation}
\label{eq:constraints}
\int \exp\left(\alpha_1+\x\trans\bb_1\right)\ps(\x|y=1)d\x=1,
\quad
\int \exp\left(\alpha_2+\x\trans\bb_2\right)\ps(\x|y=1)d\x=1.
\end{equation}

The likelihood function (\ref{eq:log-likelihood-function}) is semiparametric because we do not impose any parametric assumption on $p_s(\x|y=1)$.
Although the semiparametric likelihood enjoy benefits such as robustness, it also poses significant challenges for parameter estimation.
We cannot directly derive the maximum likelihood estimates of $(\alpha_1,\alpha_2,\bb_1,\bb_2)$ because they are intertwined with the nonparametric density $p_s(\x|y=1)$ due to the constraints in (\ref{eq:constraints}).

We propose to use the EL method (\citealt{qin1994empirical, owen2001empirical}) to estimate parameters.
The EL method allows us to estimate parameters without assuming a specific distribution assumption on $p_s(\x|y=1)$.
Specifically, we denote $p_i=\ps(\x_i|y=1)$ for $i=1,\dots,N$ and treat them as parameters.
{Then the log-EL function is given by}
\begin{equation}
\label{eq:EL}
\begin{split}
&\ell_{\text{EL}}(\alpha_1,\alpha_2,\bb_1,\bb_2,\pi,p_1,\dots,p_N)\\
=&\sum_{i=1}^{N}\log(p_i)+\sum_{j=1}^{m}\log\left\{\pi\exp\left(\alpha_1+\x\trans_{n+j}\bb_1\right)+(1-\pi)\exp\left(\alpha_2+\x\trans_{n+j}\bb_2\right)\right\}.
\end{split}
\end{equation}
{Note that the feasible values of $p_i,i=1,\dots,N$ must satisfy the following conditions}
\begin{equation}
\label{eq:EL-conditions}
p_i\geq0,\quad\sum_{i=1}^{N}p_i=1,\quad\sum_{i=1}^N p_i\exp\left(\alpha_1+\x_i\trans\bb_1\right)=1,\quad\sum_{i=1}^{N}p_i\exp\left(\alpha_2+\x_i\trans\bb_2\right)=1.
\end{equation}
The first two conditions in (\ref{eq:EL-conditions}) are due to the requirement of $\ps(\x|y=1)$ being a probability density function; the last two
conditions are from (\ref{eq:constraints}).
Now, the goal is to find out the maximum EL estimator (MELE) of unknown parameters by maximizing the log-EL function in (\ref{eq:EL}).

Let $\bt=(\alpha_1,\alpha_2,\bb_1,\bb_2,\pi)$.
Using the Lagrange multiplier method, it can be shown that the maximizer of the log-EL function for $p_i$ given $\bt$ is
\begin{equation*}
p_i=\frac{1}{N}\frac{1}
{
1+\sum_{t=1}^2\lambda_t
\left\{ \exp(\alpha_t+\x_i\trans\bb_t)-1
\right\}
},
\end{equation*}
where the Lagrange multipliers $(\lambda_1,\lambda_2)$ are solutions to 
\begin{equation}
\sum_{i=1}^N\frac{\exp(\alpha_t+\x_i\trans\bb_t)-1}
{
1+\sum_{j=1}^2\lambda_j
\left\{ \exp(\alpha_j+\x_i\trans\bb_j)-1
\right\}
}=0,\quad t=1,2.
\label{lagrange-equation}
\end{equation}
The profile log-EL function (after maximizing out $p_i,i=1,\dots,N$) of $\bt$ is given by 
\begin{eqnarray}
\ell_N(\bt)&=&
-\sum_{i=1}^N\log\left[
1+\sum_{t=1}^2\lambda_t
\left\{ \exp(\alpha_t+\x_i\trans\bb_t)-1
\right\}
\right]\nonumber\\
&&+\sum_{j=1}^{m}\log\left\{\pi\exp\left(\alpha_1+\x\trans_{n+j}\bb_1\right)+(1-\pi)\exp\left(\alpha_2+\x\trans_{n+j}\bb_2\right)\right\}.
\label{eq:optimization}
\end{eqnarray}
The MELE of $\bt$ is then defined as 
\begin{eqnarray*}
\wh{\bt}=
\operatorname*{arg\,max}_{\bt} \ell_N(\bt).
\end{eqnarray*}
The explicit form of $\wh\bt$ is generally unknown.
\update{In theory, a Newton-type algorithm could be employed to numerically calculate $\wh\bt$. However, because $\ell_N(\bt)$ is not concave, the convergence of such an algorithm is not guaranteed. Instead, the following subsection introduces an EM algorithm for numerically computing $\wh\bt$, which offers a theoretical guarantee of convergence.}

\subsection{Expectation-Maximization Algorithm} 
\label{sec:parameter_estimation}
{Due to the absence of target data labels (i.e., $\mathcal{Y}=\left\{y_{n+1},\ldots,y_{N}\right\}$), we need to deal with the mixture structure in the log-likelihood function \eqref{eq:log-likelihood-function}, 
which makes the maximization of 
\eqref{eq:optimization} difficult. 
The EM algorithm naturally fits into our problem.}

If we have access to the latent labels of the target data, the complete log-EL is
\[
\begin{split}
\ell_{\mathrm{EL}}^C(\bT)=& \update{
\sum_{i=1}^{n}\log\left\{\ps(\x_i|y=1)\right\}}\\
&\update{+
\sum_{j=1}^{m}\log\left[
\{\pi\pt(\x_{n+j}|y=1)\}^{y_{n+j}}
\{(1-\pi)\pt(\x_{n+j}|y=0)\}^{1-y_{n+j}} 
\right]}\\
=& \update{\sum_{i=1}^{n}\log\left\{\ps(\x_i|y=1)\right\}}\\
&\update{+\sum_{j=1}^{m}\left[y_{n+j}\log\left\{\pi \pt(\x_{n+j}|y=1)\right\}+(1-y_{n+j})\log\left\{(1-\pi)\pt(\x_{n+j}|y=0)\right\}\right]}\\
=&
\sum_{i=1}^{N}\log\left(p_i\right)+
\sum_{j=1}^m\left\{y_{n+j}(\alpha_1+\x_{n+j}\trans\bb_1)+(1-y_{n+j})(\alpha_2+\x_{n+j}\trans\bb_2)\right\}
\\
&+\sum_{j=1}^{m}\left\{y_{n+j}\log(\pi)+(1-y_{n+j})\log(1-\pi)\right\},
\end{split}
\]
where $\bT=(\alpha_1,\alpha_2,\bb_1,\bb_2,\pi,p_1,\dots,p_N)$ \update{and we have used the DETM \eqref{eq:detm} in the last step.} 
The EM algorithm is based on the complete log-EL.

The core of the EM algorithm is the EM iteration,
which contains an E-step and an M-step.
We use $\bT^{(r-1)}$
to denote the value  of $\bT$ after $r-1$ EM-iterations, $r=1,2,\ldots$.
When $r=1$, $\bT^{(0)}$
denotes the initial value of $\bT$.

\paragraph{E-Step} 
We need to calculate 
\[
Q(\bT|\bT^{(r-1)})=
\bbE\left\{\ell_{\mathrm{EL}}^C(\bT)|\mathcal{X}, \bT^{(r-1)}\right\},
\]
where 
$\mathcal{X}=\{\x_1,\ldots,\x_N\}$ denotes the observed features from both source and target, 
and the expectation is with respect to the conditional distribution of $\mathcal{Y}$
given $\mathcal{X}$ and using  $\bT^{(r-1)}$ for parameters of the conditional distribution.
The function $Q(\bT|\bT^{(r-1)})$ can be obtained by replacing the latent $\mathcal{Y}$ with their conditional expectations
\begin{equation*}
\begin{split}
\omega_{n+j}^{(r)}& = \bbE_t(y_{n+j} |\x_{n+j},\bT^{(r-1)})\\
&=\frac{\pi^{(r-1)}\exp(\alpha_1^{(r-1)}+\x\trans_{n+j}\bb_1^{(r-1)})}
{\pi^{(r-1)}\exp(\alpha_1^{(r-1)}+\x\trans_{n+j}\bb_1^{(r-1)})+(1-\pi^{(r-1)})\exp(\alpha_2^{(r-1)}+\x\trans_{n+j}\bb_2^{(r-1)})},
\end{split}
\label{omega-Estep}
\end{equation*}
for $j=1,\dots,m$.
It can be verified that $Q(\bT|\bT^{(r-1)})$ can be rewritten as 
\begin{equation*}
\begin{split}
Q(\bT|\bT^{(r-1)})=&\sum_{i=1}^{N}\log(p_i)+\sum_{j=1}^m\left\{\omega_{n+j}^{(r)}(\alpha_1+\x\trans_{n+j}\bb_1)+(1-\omega_{n+j}^{(r)})(\alpha_2+\x\trans_{n+j}\bb_2)\right\}\\
&+\sum_{j=1}^{m}
\left\{
\omega_{n+j}^{(r)}\log(\pi)+(1-\omega_{n+j}^{(r)})\log(1-\pi)\right\}.
\end{split}
\label{Qfun-Estep}
\end{equation*}

\paragraph{M-Step} We update $\bT$ from $\bT^{(r-1)}$ to $\bT^{(r)}$ as 
\[
\bT^{(r)}=\arg\max_{\bT}
Q(\bT|\bT^{(r-1)})\quad\mbox{subject to the constraints in (\ref{eq:EL-conditions})}.
\]
In the supplementary materials, we show that 
$\bT^{(r)}$ can be obtained in the following steps:
\begin{enumerate}
\item[]Step 1.
Update
$$
\pi^{(r)}=\frac{1}{m}\sum_{j=1}^{m}\omega_{n+j}^{(r)}.
$$
\item[]Step 2. 
Let 
\begin{equation*}
\begin{split}
Q^{(r)}(\alpha_1^\ast,\alpha_2^\ast,\bb_1,\bb_2)=&-\sum_{i=1}^{N}\log\left\{{1+\sum_{t=1}^2\exp(\alpha_t^\ast+\x_i\trans\bb_t)}\right\}\\
&+\sum_{j=1}^{m}\left\{\omega_{n+j}^{(r)}\left(\alpha_1^\ast+\x_{n+j}\trans\bb_1\right)+(1-\omega_{n+j}^{(r)})(\alpha_2^\ast+\x_{n+j}\trans\bb_2)\right\},\\
\end{split}
\end{equation*}
where
\[
\alpha_1^\ast=\alpha_1+\log\left(\frac{\sum_{j=1}^{m}\omega_{n+j}^{(r)}}{n}\right),\quad
\alpha_2^\ast=\alpha_2+\log\left(\frac{\sum_{j=1}^{m}(1-\omega_{n+j}^{(r)})}{n}\right).
\]
We update $(\alpha_1^\ast,\alpha_2^\ast,\bb_1,\bb_2)$ as 
$$
(\alpha_1^{\ast(r)},\alpha_2^{\ast(r)},\bb_1^{(r)},\bb_2^{(r)})
=\operatorname*{arg\,max}_{\alpha_1^\ast,\alpha_2^\ast,\bb_1,\bb_2}Q^{(r)}(\alpha_1^\ast,\alpha_2^\ast,\bb_1,\bb_2). 
$$
This objective function in this optimization problem is proportional to a weighted log-likelihood function of a multinomial logistic regression model with three classes.
Thus, in practice, we can easily calculate $(\alpha_1^{\ast(r)},\alpha_2^{\ast(r)},\bb_1^{(r)},\bb_2^{(r)})$ by fitting a multinomial logistic regression, which can be done by most software.

\item[]Step 3. Update $\alpha_1$ and $\alpha_2$ as 
$$
\alpha_1^{(r)}=\alpha_1^{\ast(r)}-\log\left(\frac{\sum_{j=1}^{m}\omega_{n+j}^{(r)}}{n}\right),\quad
\alpha_2^{(r)}=\alpha_2^{\ast(r)}-\log\left(\frac{\sum_{j=1}^{m}(1-\omega_{n+j}^{(r)})}{n}\right).
$$
\item[]Step 4. Update $p_{i}$ as  
\[
p_i^{(r)}=\frac{1}{n\left\{1+\exp(\alpha_1^{\ast(r)}+\x_i\trans\bb_1^{(r)})+\exp(\alpha_{2}^{\ast(r)}+\x_i\trans\bb_2^{(r)})\right\}},\quad i=1,\ldots,N.
\]
\end{enumerate}
\noindent
A summarization of the previous steps is given in the Section~S2 of the supplementary materials.

The following proposition shows that
log-EL 
$
\ell_{\text{EL}}(\bT)
\equiv
\ell_{\text{EL}}(\alpha_1,\alpha_2,\bb_1,\bb_2,\pi,p_1,\dots,p_N)$ does not  decrease after each iteration.

\begin{proposition}
\label{theorem.em}
With the EM algorithm described above, we have for {$r\geq1$}
$$
\ell_{\text{EL}}(\bT^{(r)})
\geq
\ell_{\text{EL}}(\bT^{(r-1)}).
$$
\end{proposition}

The proof of Proposition \ref{theorem.em} is given in the supplementary materials.
We make two remarks about the EM algorithm. First,
note that
\begin{equation*}
\ell_{\text{EL}}(\bT)=\sum_{i=1}^{n}\log(p_i)+\sum_{j=1}^{m}\log\left\{ \pi p_{n+j}\exp\left(\alpha_1+\x\trans_{n+j}\bb_1\right)+(1-\pi)p_{n+j}\exp\left(\alpha_2+\x\trans_{n+j}\bb_2\right)\right\},
\end{equation*}
which, together with \eqref{eq:EL-conditions}, implies that 
{$\ell_{\text{EL}}(\bT)\leq 0$}. 
With this result, Proposition \ref{theorem.em} guarantees that the EM algorithm eventually converges
to at least a local maximum for the given initial value $\bT^{(0)}$.
\update{
Practically, we recommend using multiple initial values to explore the likelihood function to ensure achieving the global maximum.}
In addition, one can stop the algorithm when the increment in the log-EL after an iteration is no greater than, say,
1e-4. 

\section{\label{sec:theory}Theoretical Results}

\subsection{Statistical Inference}
\label{subsec:inference}
In this section, we first provide the asymptotic distribution of the MELE $\wh\bt$.
Then, we focus on two specific statistical inference tasks: the first one is the goodness-of-fit test of the SETM as opposed to the DETM; the second one is the inference on the mixture proportion $\pi$.

We need some notation for developing the asymptotic distribution for $\wh\bt$.
Let $(\wh\lambda_1,\wh\lambda_2)$ be the Lagrange multipliers corresponding to $\wh\bt$, i.e., 
$(\wh\lambda_1,\wh\lambda_2)$ is the solution of \eqref{lagrange-equation} with $\bt$ being replaced by $\wh\bt$.
 Throughout the paper, we assume that the total sample size $N = n+m \to \infty$ and $m/N\to c$ for some constant $c \in(0,1)$.
This assumption indicates that $n$ and $m$ go to infinity at the same rate. 
For simplicity and presentation convenience, we write $c = m/N$ and assume it is a constant, which does not affect our technical development. 
 The following theorem establishes the joint asymptotic normality of $(\wh\lambda_1,\wh\lambda_2)$ and $\wh\bt$, which implies the asymptotic normality of $\wh\bt$.
The proofs for results in this section are given in the {supplementary materials}.

\begin{theorem}
		\label{th__asm_normality}
	Suppose the DETM and conditions in Section S4.1 of the supplementary materials
 are satisfied.  
 Let $\bt^{o}=(\alpha_1^o,\alpha_2^o,\bb_1^o,\bb_2^o,\pi^o)$ be the true value of $\bt$.
As $N$ goes to infinity, 
  $$
  \sqrt{N}\left\{
    \widehat\lambda_1-c\pi^{o},
  \widehat\lambda_2-c(1-\pi^{o}), 
    (\wh\bt-\bt^{o})\trans
  \right\}\trans  \rightarrow N\left(\0, \bSigma\right)
  $$ in distribution,  where $\bSigma$ is given in Equation (14) in the supplementary materials. 
	\end{theorem}

For CD PU data, using the DETM might not be the best approach as SETM could improve computational and estimation efficiency in such cases.
To help us decide whether the SETM can provide a good fit to the PU data, we can use the profile log-EL function in \eqref{eq:optimization} to construct the goodness-of-fit test of the SETM as opposed to the DETM. 
Specifically, the empirical likelihood ratio (ELR) test for testing SETM vs. DETM can be formulated as the following null hypothesis
$H_0:\alpha_1=0$ and $\bb_1=\0$ with test statistic given by
$$
R_{N}
=2\left\{\ell_N(\wh\bt)
-\ell_N(\wt\bt)
\right\},
$$
where $\wt\bt$ is the MELE of $\bt$ under the SETM with $\alpha_1=0$ and $\bb_1={\bf 0}$, and $\wh\bt$ is the MELE under the DETM.
Note that $\wt\bt$ can be obtained by slightly modifying the EM algorithm: we only need to fix $\alpha_1=0$ and $\bb_1={\bf 0}$ in the E-step and  M-step. 

For the inference of $\pi$, the profile log-EL function in \eqref{eq:optimization} can also be used to define the ELR function of $\pi$ as
$$
R_N^{\ast}(\pi)=2\left\{\ell_N(\wh\bt)
-\ell_N(\wh\bt_{\pi})
\right\},
$$
where $\wh\bt_{\pi}$ is the MELE of $\bt$ with $\pi$ being fixed. 
We can derive $\wh\bt_{\pi}$ as if we know the value of $\pi$ in the EM algorithm: we only need to fix $\pi$ in the E-step and M-step.

Let $\pi^{o}$ be the true value of $\pi$ with $0<\pi^{o}<1$.
The following theorem summarizes the asymptotic results of $R_{N}$ and $R_N^{\ast}(\pi^{o})$.
\begin{theorem}
    \label{prop_pi_bb_1}
   Assume the same conditions as in Theorem \ref{th__asm_normality}.  We have
    \begin{itemize}
        \item[(a)] $R_{N}\xrightarrow{d}\chi_p^2$ as $N\to\infty$ under the null hypothesis $H_0:\alpha_1=0$ and $\bb_1={\bf 0}$, where $p$ is the dimension of $\bb_1$.
        \item[(b)]
        $R_N^{\ast}(\pi^{o})\xrightarrow{d}\chi_1^2$ as $N\to\infty$.
    \end{itemize}
\end{theorem}
Based on Theorem~\ref{prop_pi_bb_1}, we can perform hypothesis tests for $\pi$ and the goodness-of-fit of the SETM on the PU data.
For example, we can reject the SETM and adopt the DETM at significant level $0.05$ if $R_N>\chi^2_{p,0.95}$, where $\chi^2_{p,0.95}$ is the 95\% quantile of $\chi^2_p$.
In addition, we can also construct the ELR-based confidence interval for the mixture proportion $\pi$.
For example, a $95\%$ confidence interval for $\pi$ is given by $\left\{\pi:R_N^\ast(\pi)\leq\chi^2_{1,0.95}\right\}$.

\subsection{Approximation of the Bayes Classifier}
Recall that the Bayes classifier in Section \ref{subsec:motivation-det} depends on ${\Pr}_{t}(Y=1|\X=\x)$ in \eqref{eq:bayes-posterior}.
Under the DETM in \eqref{eq:detm}, ${\Pr}_{t}(Y=1|\X=\x)$ can be rewritten as 
\begin{equation*}
\phi(\x;\bt)\equiv
{\Pr}_{t}(Y=1|\X=\x)=
\frac{\pi\exp\left(\alpha_1+\x\trans\bb_1\right)}{\pi\exp\left(\alpha_1+\x\trans\bb_1\right)+(1-\pi)\exp\left(\alpha_2+\x\trans\bb_2\right)}.
\end{equation*}
With the MELE of $\bt$, we estimate ${\Pr}_{t}(Y=1|\X=\x)$ by $\phi(\x;\wh\bt)$, thus approximating the Bayes classifier.
In the following theorem, we show that the $L_1$-distance between 
$\phi(\x;\wh\bt)$ and $\phi(\x;\bt^o)$ has the order $N^{-1/2}$. 

\begin{theorem}
\label{theorem4:bayes}
Assume the same conditions as in Theorem  \ref{th__asm_normality}.
We have 
$$
\int \left|\phi(\x;\wh\bt)-\phi(\x;\bt^o)\right|p_t(\x)d\x=O_p(N^{-1/2}). 
$$

\end{theorem}

\section{Numerical Studies}
\label{sec:numerical}
In this section, we first present a simulation study to evaluate the proposed methods.
Then, we apply our method to a real data application to predict the selling price of mobile phones.
We use our custom R package \texttt{PUEM} in this section, the package is available {in the supplementary materials.}
Some benchmark methods were implemented in {Python} because the authors only provide Python source code.

\subsection{\update{Simulation Study}}
We conduct a simulation study to assess the performance of our proposed framework across various tasks.
We are interested in the following tasks: 
1) Conducting a goodness-of-fit test between SETM and DETM, focusing on the type-I error and power of the test procedure.
2) Evaluating the root mean square error (RMSE) of the MELE of $\pi$.
3) 
Determining the coverage accuracy of the confidence intervals of $\pi$.
4) 
Assessing the classification accuracy using the approximated Bayes classifier.

\paragraph{Detection of CD Data}
Let $\0_p$ and $\1_p$ denote the p-dimensional column vectors of zeros and ones, respectively. We generate data from the following multivariate normal distributions:
1) Source positive distribution: $p_s(\x|y=1)\sim\mathrm{MVN}(\bmu_1,\mathbf{I})$, where $\bmu_1=\0_{15}\in\mathbb{R}^{15}$ and $\I$ is the identity matrix.
2) Target negative distribution: $p_t(\x|y=0)\sim\mathrm{MVN}(\bmu_2,\mathbf{I})$, where $\bmu_2=\1_{15}\in\mathbb{R}^{15}$.
3) Target positive distribution: $p_t(\x|y=1)\sim\mathrm{MVN}(\bmu_3,\mathbf{I})$, where we can choose different levels for $\bmu_3\in\left\{\0_{15}, (1 ~\0_{14}\trans)\trans,\dots,(\1_{7}\trans ~\0_{8}\trans)\trans\right\}.$
It should be noted that when $\bmu_3=\0_{15}$, the PU data are CD; otherwise, the data are DD.
We perform the goodness-of-fit test for SETM vs. DETM, where the null hypothesis is that the PU data are CD and follow the SETM, which can be translated to $H_0:\alpha_1=0\mbox{ and } \bb_1=\0_{15}$.

In the experiment, we set the sample size of the source data equal to that of the target data ($n=m$) and fix the proportion of positive data on the target at 75\% ($\pi=0.75$).
The levels of $n$ are $1000, 2000, 3000, 4000,$ and $5000$.
Each factor level combination is repeated {{1000}} times.
Table \ref{tab:goodness-of-fit} summarizes the rejection rates of the goodness-of-fit test at the 5\% significance level.

\begin{table}[!ht]
\centering
\caption{Rejection rates of the goodness-of-fit test for testing $H_0:\alpha_1=0\mbox{ and } \bb_1=\0_{15}$.}
\begin{tabular}{crrrrrrrr}
\toprule
\multirow{2}{*}{$n$}        & \multicolumn{8}{c}{Number of ``1''s in $\bmu_3$}                                                                                                                                                  \\ \cline{2-9} 
                          & \multicolumn{1}{c}{0} & \multicolumn{1}{c}{1} & \multicolumn{1}{c}{2} & \multicolumn{1}{c}{3} & \multicolumn{1}{c}{4} & \multicolumn{1}{c}{5} & \multicolumn{1}{c}{6} & \multicolumn{1}{c}{7} \\ \hline
\multicolumn{1}{c}{1000} & 10.7\%                & 100\%                 & 100\%                 & 100\%                 & 100\%                 & 100\%                 & 100\%                 & 100\%                \\
\multicolumn{1}{c}{2000} & 7.4\%                 & 100\%                 & 100\%                 & 100\%                 & 100\%                 & 100\%                 & 100\%                 & 100\%                 \\
\multicolumn{1}{c}{3000} & 5.3\%                 & 100\%                 & 100\%                 & 100\%                 & 100\%                 & 100\%                 & 100\%                 & 100\%                 \\
\multicolumn{1}{c}{4000} & 5.6\%                 & 100\%                 & 100\%                 & 100\%                 & 100\%                 & 100\%                 & 100\%                 & 100\%                 \\
\multicolumn{1}{c}{5000} & 5.0\%                 & 100\%                 & 100\%                 & 100\%                 & 100\%                 & 100\%                 & 100\%                 & 100\%                 \\ \bottomrule
\end{tabular}
\label{tab:goodness-of-fit}
\end{table}
When $\bmu_1=\bmu_3=\0_{15}$ (i.e., the null hypothesis is true; see column ``0''), the rejection rate (i.e., the type-I error) converges to the nominal significant value $0.05$ as sample size $n$ increases, which honors the asymptotic results in Section \ref{subsec:inference}.
For columns ``1'' through ``7'', the null hypothesis does not hold because $\bmu_3\neq\bmu_1$.
To our surprise, the rejection rates (i.e., power) are uniformly 100\%, indicating that the proposed hypothesis test will always reject the null, even if $\bmu_3$ and $\bmu_1$ differs in only one dimension.

\paragraph{Estimations and Confidence Intervals for the Mixture Proportion}
We use two different data-generating mechanisms in this task.
The first one generates CD PU data, where the source positive (i.e., $p_s(\x|y=1)$) and target positive (i.e., $p_t(\x|y=1)$) distributions are the same and are set as $\mathrm{MVN}(\0_{15},\mathbf{I})$; whereas the target negative distribution is set as $p_t(\x|y=0)\sim\mathrm{MVN}(\1_{15},\mathbf{I})$.
The second one generates DD PU data, where the source positive distribution is set as $\mathrm{MVN}(\0_{15},\mathbf{I})$, the target positive distribution is set as $\mathrm{MVN}((\1\trans_{7}~\0_{8}\trans)\trans,\mathbf{I})$, and the target negative distribution is $\mathrm{MVN}(\1_{15},\mathbf{I})$.

Similar to the previous task, we let $n=m$ and vary $n$ in $\{1000, 2000, 3000, 4000, 5000\}$.
We use two values of $\pi\in\{0.3, 0.7\}$ and repeat each factor level combination for 1000 times.
We estimate the mixture proportion $\pi=\Pr_t(Y=1)$ and construct confidence intervals using both {the} DETM and SETM.

\update{
The estimation results are summarized in Table \ref{tab:pi-estimation_bias_RMSE}.
In the case of the CD PU data, both the DETM and SETM provide good estimates.
In the left panel of Table~\ref{tab:pi-estimation_bias_RMSE}, we see that the estimated bias, and RMSE are all small.
Regarding the estimation efficiency, we observe that the SETM outperforms DETM, albeit marginally.
The efficiency gain is due to that the SETM is a reduced model of DETM.
For the DD data, the DETM method still performs well with RMSE shrinking toward 0 as the sample size increases.
However, the SETM fails due to model misspecification as illustrated in the right panel of Table~\ref{tab:pi-estimation_bias_RMSE}.
}

\begin{table}[!ht]
\centering
\caption{Sample means and sample root mean square error (RMSE, in parenthesis) of $\pi$ estimates.}
\label{tab:pi-estimation_bias_RMSE}
\begin{adjustbox}{max width=\textwidth}
\begin{tabular}{@{}crlllrrrr@{}}
\toprule
\multirow{3}{*}{$n$} & \multicolumn{4}{c}{CD Data}                                                                              & \multicolumn{4}{c}{DD Data}                                                                               \\ \cmidrule(l){2-5}\cmidrule(l){6-9} 
                   & \multicolumn{2}{c}{SETM}                    & \multicolumn{2}{c}{DETM}                     & \multicolumn{2}{c}{SETM}                    & \multicolumn{2}{c}{DETM}                     \\ \cmidrule(l){2-5}\cmidrule(l){6-9} 
                   & \multicolumn{1}{c}{$\pi=0.3$} & \multicolumn{1}{c}{$\pi=0.7$} & \multicolumn{1}{c}{$\pi=0.3$} & \multicolumn{1}{c}{$\pi=0.7$} & \multicolumn{1}{c}{$\pi=0.3$} & \multicolumn{1}{c}{$\pi=0.7$} & \multicolumn{1}{c}{$\pi=0.3$} & \multicolumn{1}{c}{$\pi=0.7$} \\ \cmidrule(l){2-5}\cmidrule(l){6-9} 
1000               & 0.299 (0.016)           & 0.701(0.019)
& 0.300 (0.019)           &   0.703(0.021)       &  0.001(0.299)          &   0.001(0.699)       &    0.301(0.070)         &    0.672(0.078)       \\
2000               & 0.300 (0.012)           &  0.701(0.013)          & 0.300 (0.013)           &  0.701(0.014)         &  0.001(0.299)        &  0.001(0.699)           &    0.300(0.042)        &   0.687(0.053)        \\
3000               & 0.300 (0.009)           & 0.700(0.010)
& 0.300 (0.010)           &   0.700(0.012)         &   0.001(0.299)        &    0.001(0.699)       &   0.300(0.031)           &  0.687(0.044)       \\
4000               & 0.300 (0.008)           &  0.700(0.009)         & 0.300 (0.009)           &  0.700(0.010)    &   0.001(0.299)      &    0.000(0.700)        &   0.299(0.026)       &  0.691(0.033)        \\
5000               & 0.300 (0.007)            &     0.700(0.008)         
& 0.300 (0.008)           &    0.700(0.009)       &  0.001(0.299)          &  0.000(0.700)          &   0.300(0.023)         &  0.694(0.028)       \\ \bottomrule
\end{tabular}
\end{adjustbox}
\end{table}

\begin{figure}[ht!]
    \centering    \includegraphics[width=0.9\linewidth]{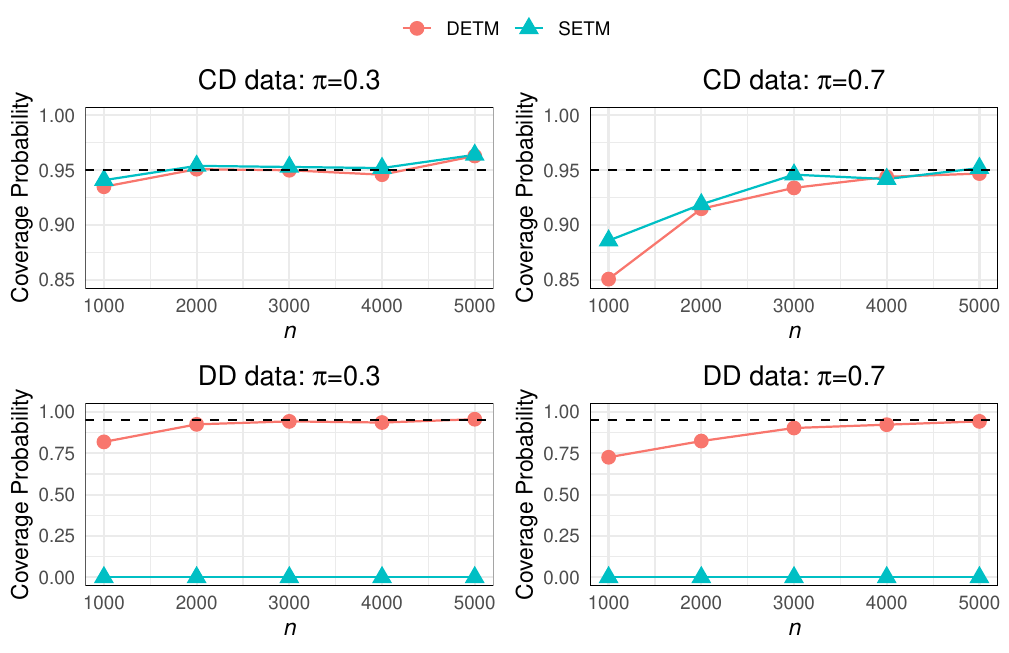}
    \caption{Coverage probabilities of 95\% confidence intervals for $\pi$.}
    \label{fig:coverage}
\end{figure}

We also construct two-sided $95\%$ confidence intervals for $\pi$, and the results are given in Figure~\ref{fig:coverage}.
We report the estimated coverage probabilities in each of the different scenarios: 1) SETM on the CD PU data, 2) DETM on the CD PU data, 3) SETM on the DD PU data, and 4) DETM on the DD PU data.
For the CD data, the coverage probabilities of both DETM and SETM reach the nominal level of 95\% as the sample size increases.
However, when the data are DD, the confidence intervals constructed using the SETM fail to cover the true parameter value $\pi$, resulting in a coverage probability that is close to 0.
On the other hand, the DETM's confidence intervals still perform well, remaining around the nominal levels of 95\%.

\paragraph{Classification Accuracy}
Lastly, we assess the prediction performance of our proposed method.
{We generate the CD and DD data the same way as in the ``Estimations and Confidence Intervals for the Mixture Proportion'' part, where $\pi\in \{0.3, 0.7\}$ and $n=m\in\{1000, 2000, 3000, 4000, 5000\}$.}
To avoid overfitting, we generate a new validation dataset from the target distribution $p_t(\x,y)$ with sample size $n$, and the classifiers are applied to the validation dataset to calculate the classification accuracy.

We compare the following methods, including ours and some existing methods in the machine learning literature.
1) DETM: the proposed approximated Bayes classifier by replacing  parameters with the MELEs under the DETM.
2) SETM: the approximated Bayes classifier by replacing parameters with the MELEs under the SETM.
3) PS: proposed by \citet{bekker2020beyond} and is based on the propensity score function.
4) KM2: proposed by \citet{ramaswamy_mixture_2016} using the reproducing kernel Hilbert space and is based on CD PU data.
5) TICE: proposed by \citet{bekker2018tice} and is based on CD PU data.

\begin{figure}[ht]
    \centering    \includegraphics[width=0.9\linewidth]{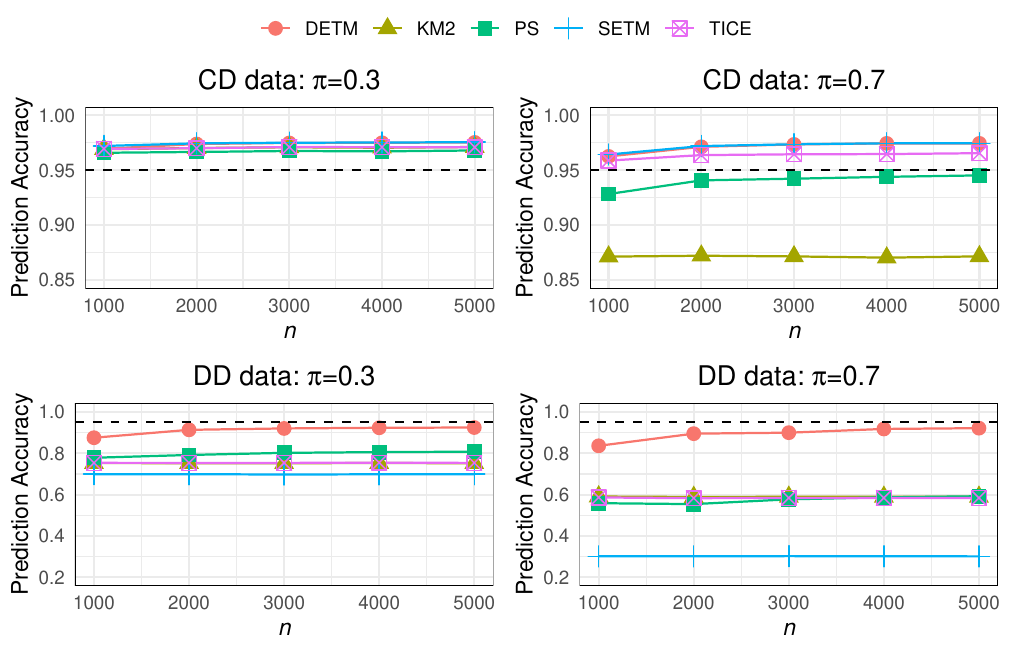}
    \caption{Classification accuracies of five classifiers.}
    \label{fig:prediction}
\end{figure}

Figure~\ref{fig:prediction} illustrates the classification accuracies using 100 replicates.
When the data are CD and $\pi=0.3$, all classifiers perform well, achieving an accuracy exceeding 95\% in all sample sizes.
However, when $\pi=0.7$, the KM2 method has the lowest accuracy at around 87\%, while all other methods maintain at least 92\% accuracy in all sample sizes.

For the DD data, the simulation studies highlight discrepancies among different methods. In both $\pi=0.3$ and $\pi=0.7$ scenarios, our proposed DETM method achieves the highest classification rate of around 90\%, with accuracy improving as the sample size increases.
In comparison, at $\pi=0.3$, all benchmark methods have accuracies between 75\% and 81\%, with SETM performing the worst across all sample sizes.
When $\pi=0.7$, the accuracies of benchmark methods PS, KM2, and TICE are around 60\%.
SETM again has the worst performance.

\paragraph{Summary}
The simulation study demonstrates that our proposed DETM method excels in estimation, inference, and classification tasks, regardless of whether DD or CD data are used. Under correct model assumptions, the SETM method performs well, with slightly improved estimation efficiency compared to DETM. However, SETM lacks robustness and performs poorly across all tasks when the CD assumption is violated. Additionally, the goodness-of-fit test effectively distinguishes DD data by accurately identifying CD data with a type-I error rate close to the nominal level and exhibiting high power in detecting non-CD data.
For classification tasks, the proposed DETM method also shows superior performance compared to existing methods.
\update{We also conduct additional simulations for extreme scenarios, such as when \(\pi\) is close to 0. However, due to space limitations, these results are included in the supplementary materials.
}

\subsection{Mobile Phone Dataset}

This section demonstrates the proposed methodology using a real data application.
We consider the Mobile Phone Dataset from the Kaggle \footnote{Available at \href{https://www.kaggle.com/datasets/iabhishekofficial/mobile-price-classification}{https://www.kaggle.com/datasets/iabhishekofficial/mobile-price-classification}}.
The Mobile Phone Dataset contains 20 features and an ordinal label taking values in the set $\{0, 1, 2, 3\}$ that indicates the phone's price range from low cost to very high cost.
Under each class, there are 500 observations.
The features include properties such as the memory size and the phone's weight; see Table~S3 in the supplementary materials for the full list of the features.

Before the analysis, we first preprocess the dataset to fit our needs.
We first merge classes 0 and 1 to form the  target positive data from $p_t(\x|y=1)$ (low-end phone).
We then use class 2 as the source positive data from $p_s(\x|y=1)$ (low-end phone) and class 3 as target negative data from $p_t(\x|y=0)$ (high-end phone).
Then, the source positive data have a sample size of $n=500$, and the target data have a sample size of $m=1500$, where $1000$ are positive and $500$ are negative.
Our formulation ensures that the low-end phones in the source and target are different, as low-end phones in the target data are less expensive; thus, the data can be seen as DD PU.

\begin{figure}[ht!]
    \centering    \includegraphics[width=0.7\linewidth]{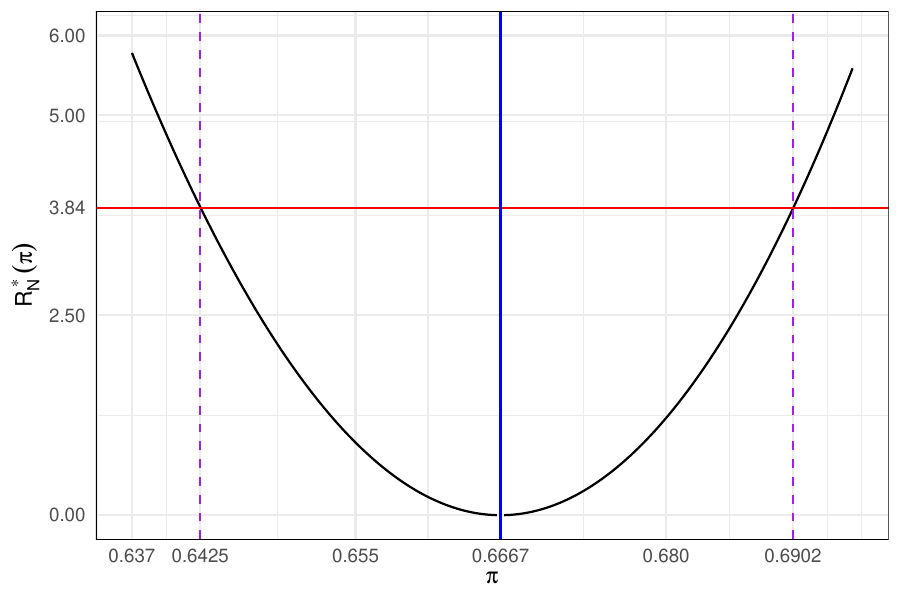}
    \caption{Plot of the ELR function $R_N^{\ast}(\pi)$ versus $\pi$.}
    \label{fig:app-llk_pi}
\end{figure}

We first {conduct} a goodness-of-fit test for SETM vs. DETM.
The value of the ELR statistic is $R_N=1302.222$, which gives a p-value of 0, calibrated by the limiting distribution $\chi^2_{20}$.
Hence, we reject the null hypothesis with overwhelming evidence and adopt the DETM.
The test results also align with the way we construct the PU data.
Next, we {examine} the estimation and inference of the mixture proportion $\pi$.
The true $\pi$ in the target data is $\pi^{o}=2/3$, 
and the estimate $\widehat\pi=0.6667$ is close to the true value.
For the inference of $\pi$, we examine the behavior of the ELR function $R_N^{\ast}(\pi)$ in Figure \ref{fig:app-llk_pi}.
The vertical solid blue line is the true value of $\pi$.
The red horizontal line is the 95\% quantile of $\chi^2_1$, and the two dashed lines indicate the two endpoints of a $95\%$ confidence interval.
We can see that the true value is within the confidence interval $[0.6425, 0.6902]$.

Lastly, we focus on the classification.
Recall that the sample size of the positive target, positive source, and negative target data are $n_1=1000, n_2=500$, and $n_3=500$, respectively.
To avoid overfitting, we reserve 20\% of the positive target and negative target data and combine them for verification.
Thus, the PU data have a sample size of $N=n+m$, where $n=n_2=500$ and $m=0.8n_1+0.8n_3=1200$.
The sample size of the verification dataset is $0.2n_1+0.2n_3$.
We repeated the data partition on the target data for 20 times and performed classification for each replicate.
We compared the proposed DETM method with the benchmarks used in the simulation study.
Table \ref{tab:app-prediction} summarizes the classification accuracy averaged from $20$ sample splitting replicates.
The numbers in the parenthesis are the standard deviation of the 20 accuracies from 20 replicates.
Our proposed method, DETM, has surprisingly achieved perfect classification, while the benchmark methods have accuracies of about 88. 20\% to 91. 30\%.

\begin{table}[h!]
\caption{Classification accuracies on the Mobile Phone Dataset.}
\label{tab:app-prediction}
\centering
\begin{adjustbox}{max width=\textwidth}
\begin{tabular}{ccccccc}
\hline
Method & DETM  & PS & KM2 & TICE \\ \hline
Accuracy & 100\% (0\%)&   91.30\% (1.39\%)        & 89.05\% (3.31\%)       & 88.20\% (1.46\%)           \\ \hline
\end{tabular}
\end{adjustbox}
\end{table}


\section{Concluding Remarks}\label{sec-conclu}

{
\color{black}
We would like to compare our transfer learning framework with the propensity score framework used in the literature (\citealt{bekker2020beyond,bekker_learning_2020}) in this final section.
Unlike the transfer learning setting, which involves two samples, the propensity score framework considers only a single sample, where both labeled and unlabeled data are assumed to be generated from a joint distribution $p(\x,y)$.
 In this framework, an additional binary random variable $S$ is introduced to determine whether an example is labeled ($S=1$) or unlabeled ($S=0$).
The conditional distribution for positive examples, $\Pr(S=1|\x,y=1)$, is called the propensity score function. 
It is  unnecessary to consider $y=0$, as 
$\Pr(S=1|\x,y=0)=0$.
If the propensity score function does not depend on $\x$ (i.e., $\Pr(S=1|\x,y=1)$ is a constant), we say the PU data are selected completely at random (SCAR).
Otherwise, the PU data are selected at random (SAR).
}

\blue{The SCAR and SAR PU settings correspond to the CD and DD assumptions, respectively.
To illustrate this, we can write the propensity score function as
\begin{equation}
\label{eq:propensity_score_and transfer_learning}
\begin{split}
&\Pr(S=1|\x,y=1)=\frac{p(S=1,\x,y=1)}{p(\x,y=1)}\\
=&\frac{p(\x|y=1,S=1)\Pr(S=1,y=1)}{p(\x|y=1,S=0)\Pr(S=0,y=1)+p(\x|y=1,S=1)\Pr(S=1,y=1)}\\
=&\frac{\Pr(S=1,y=1)}{\dfrac{p(\x|y=1,S=0)}{p(\x|y=1,S=1)}\Pr(S=0,y=1)+\Pr(S=1,y=1)}.
\end{split}
\end{equation}
Notice that $p_s(\x|y=1)=p(\x|y=1,S=1)$ and $p_t(\x|y=1)=p(\x|y=1,S=0)$.
Then from (\ref{eq:propensity_score_and transfer_learning}) we can see that the CD assumption is equivalent to the SCAR assumption.
Also, when the propensity score does depend on $\x$ (i.e., SAR PU), it implies that the CD assumption does not hold.
In other words, the SAR PU setting implies the DD PU setting under the transfer learning framework.
}

\blue{
What does the DETM imply about the propensity score function?
From (\ref{eq:propensity_score_and transfer_learning}), we can see that under the DETM, the propensity score function follows a logistic regression model:
\[
\Pr(S=1|\x,y=1)=\frac{\exp(\alpha^\prime_1+\x\trans\bb_1)}{1+\exp(\alpha^\prime_1+\x\trans\bb_1)}~~
\mbox{with}~~
\alpha^\prime_1=\alpha_1+\log\left\{\frac{\Pr(S=1,y=1)}{\Pr(S=0,y=1)}\right\}.
\]}

We conclude this paper by highlighting several promising directions for future research. One intriguing extension involves exploring the open set label shift setting (e.g., \citealt{garg_domain_2022}), where the target domain introduces a new class not present in the source domain, but the source domain has multiple classes.
This setting expands the scope of PU data and presents new challenges and opportunities for DETM to demonstrate its adaptability and effectiveness.
Another critical area for future investigation is the robustness of DETM in the presence of noisy labels. 
Enhancing the model's resilience to such imperfections is vital for ensuring reliable performance in real-world applications, where data quality can often be compromised. \update{Furthermore, if the dimension of $\bX$ is high, penalized methods may be considered before fitting the ETMs. Through the estimation procedures presented in Section~\ref{sec:parameter_estimation}, particularly Step 2 in the M-step, we may, for instance, use \texttt{glmnet} to easily incorporate penalized maximum (empirical likelihood) methods to deal with high-dimensional $\bX$.}

\setcounter{section}{1}

\section*{Supplementary Materials}
The online supplementary document includes all technical details and proofs, additional details on the M-step of the EM algorithm, form of $\bSigma$ and conditions, extra simulation studies, and features of the mobile phone dataset. We also provide the R package and all necessary code to reproduce the analysis.

\section*{Acknowledgment}

We sincerely thank the joint editor and associate editor for their time and effort in overseeing the review of this manuscript. We are also grateful to the three anonymous reviewers for their insightful comments, which greatly improved our work. Additionally, we appreciate Dr. Xiyue Han for his helpful discussions in the early stages of this research. Finally, we extend our gratitude to Professor Yukun Liu for his valuable and constructive feedback.
This work is partially supported by the Natural Sciences and Engineering Research Council of Canada (RGPIN-2023-03479 and RGPIN-2020-04964) and the Faculty of Mathematics Research Chair Funding of the University of Waterloo.

The majority of the work was completed during Siyan Liu's visit to the University of Waterloo as a visiting PhD student under the joint supervision of Qinglong Tian and Pengfei Li. Qinglong Tian is the corresponding author.

\section*{Disclosure Statement}
No potential conflict of interest was reported by the authors. 

\bibliographystyle{apalike}
\bibliography{ref}

\end{document}